\documentclass[12pt,onecolumn,amsmath,amssymb]{revtex4}
\usepackage{graphicx} 
\usepackage{hyperref} 
\usepackage{bm}
\usepackage{color}
\begin{document}

\title{Boundary induced phase transition with stochastic entrance and exit}

\author{Mithun Kumar Mitra$^1$ and Sakuntala Chatterjee$^2$ }

\affiliation{$^1$ Department of Physics, Indian Institute of Technology - Bombay, Powai, Mumbai 400076, India 
\\ $^2$ Department of Theoretical Sciences, S. N. Bose National Centre for Basic Sciences, Block - JD, Sector - III, Salt Lake, Kolkata 700098, India }

\begin{abstract}
Abstract: We study an open-chain totally asymmetric exclusion process (TASEP) 
with stochastic gates present at the two boundaries. The gating dynamics has
been modeled keeping the physical system of ion-channel gating in mind.
These gates can randomly
switch between an open state and a closed state. In the open state, the gates
are highly permeable such that any particle arriving at the gate immediately 
passes through.  In the closed state, a particle gets trapped at the gate and 
cannot pass through until the gate switches open again. We calculate the
phase-diagram of the system and find important and non-trivial differences with
the phase-diagram of a regular open-chain TASEP. In particular, depending on 
switching rates of the two gates, the system may or may not admit a maximal
current phase. Our analytic calculation within mean-field theory captures 
the main qualitative features of our Monte Carlo simulation results.  
We also perform a refined mean-field calculation where the correlations at the 
boundaries are taken into account. This theory shows significantly better 
quantitative agreement with our simulation results.
\end{abstract}

\maketitle

\section{Introduction}

The totally asymmetric simple exclusion process (TASEP) is a paradigmatic
 model for driven diffusive systems \cite{gunter}.
 TASEP is particularly used for modeling transport in
a wide variety of nonequilibrium systems, {\sl e.g.} motion of molecular
motors on microtubules  \cite{menon,lipowsky,klumpp},
 filamentous fungal growth \cite{evans1,evans2,evans3} and flow of traffic 
on a road \cite{schreck,schads,helbing,hilhorst}.
The model is defined on a one dimensional lattice on which hardcore
particles perform biased diffusion. Despite its simplicity, this model shows
interesting properties like the existence of a phase transition in one
 dimension, which are hallmarks of nonequilibrium systems \cite{krug,henkel}.

One widely-studied variant of the TASEP is defined on an open lattice where in
addition to performing biased diffusion, the system can also exchange particles
 with the reservoirs at the boundaries. In this open system, particles
are injected at one end and extracted at the other end of the lattice and by
tuning these injection and extraction rates one finds that the system shows
a boundary induced phase transition \cite{domany,derrida,gunter93}. 
The resulting phase diagram was derived using exact calculations and
an alternative simpler derivation was also presented using mean field theory
\cite{domany}. The phase diagram consists of a high density phase, a low
density phase and a maximal current phase. As the names suggest, the high 
density (low density) phase is characterized by a large (small) bulk density 
and the maximal current phase by the presence of largest admissible current in 
the system.

The open chain TASEP is widely used to model transport in complex biological
systems \cite{zia,chowdhury}. Owing to the inherent complexity of a biological 
system, transport processes are generally coupled with other biochemical
 reactions. At the entry and exit terminals other reactions can take
place whose outcome may influence the injection and exit processes strongly. 
Such a situation was considered in \cite{wood} where the species of particles
undergoing transport were assumed to take part in other reactions before or
after the transport {\sl i.e.} entrance or exit are possible only when certain
reactions take place. The system was modeled as an open-chain TASEP with
stochastically mediated entrance or exit, facilitated by gates present at the
boundaries. The opening and closing of the gate at the exit (entrance)
 terminal is determined by the  binding and unbinding of certain receptor
 (initiator) to the gate.  When the
  receptor (initiator) is bound to the gate, the particles can pass through
the gate and exit (enter) the channel. When the gate at the exit (entrance)
end is not bound to any receptor (initiator), it is assumed to be in a closed
state when no particle can pass through. One important component in this model
is the two-way coupling between the gate dynamics and the particle flow. When
a particle passes through an open gate, the receptor (initiator) immediately
unbinds, {\sl i.e.} the particles `close the gate behind them'. The new phase
diagram for this model was studied using Monte Carlo simulations and
mean-field theory.

One physical system, where stochastic gates are explicitly present and
control cellular transport, can be found in ion-channels, which are embedded
 on cell membrane and  exchange of ions between cells takes place through these
channels \cite{hille}.
 An ion-channel stochastically switches between  `open' and `closed' 
conformations and while the open conformation allows fast passage of ions
through the channel resulting in a large ion-flux, no transport is possible in
the closed state \cite{hodgkin,nehar}.
 The closed state is also known as the high-affinity state,
when the ions strongly bind to the channel and hence gets trapped.
In the open conformation, the channel is in the low-affinity state
and the ions immediately dissociate from the channel and move to the cell 
cytosol.

In this paper, we explicitly model the gating dynamics, following the
modeling strategy used to describe ion-channel gating
\cite{dongen, andre11}, and  study its 
effect on the transport. We consider a generalization of an open chain TASEP,  
where the injection and extraction processes at the two boundaries are
 controlled
by the presence of gates or special boundary sites, which can stochastically
switch between high- and low-affinity states, as described above. While
in the high-affinity state, a particle gets trapped in the gate, in the
low-affinity state it immediately  dissociates from the gate and moves towards
right.

Our Monte Carlo simulations show 
that the inclusion of the above gating mechanism produces interesting
and non-trivial modifications in the phase diagram. The residence time of the
gates in the open state are important parameters here and depending on their
value all three phases may or may not be observed. For example, 
unlike in ordinary TASEP,
sufficiently large values of the injection and exit rates do not always yield
an MC phase --- the residence time of the gates in the open state must exceed
certain critical values for MC phase to be observed. To explain these results
we perform analytical calculations within mean-field theory. A simple
mean-field theory where all correlations in the system are ignored, shows
qualitative agreement with the simulation results but there is significant
quantitative mismatch. To improve the quantitative agreement, we use a 
refined mean-field theory, developed in \cite{evans3}, where important
correlations at the boundaries are taken into
account and this approach yields better agreement with the simulation data.

In the next section we briefly summarize earlier results on the phase diagram 
for an ordinary open chain TASEP and then introduce our stochastic gated model 
in detail. In section \ref{sec:sim}, we present the results of Monte Carlo 
simulation of our model. Our analytical calculations within simple and refined
mean-field theory are presented in sections \ref{sec:mft} and \ref{sec:rmft},
respectively. Discussion and conclusions are presented in 
section \ref{sec:discussion}.

\section{Description of the Model}
\label{sec:model}

The usual open-chain TASEP is defined on a one dimensional open lattice of
$L$ sites. At each site there can be at most one particle which can hop towards
  right, if the right neighboring site is empty. Particles are injected and
extracted at the two ends of the lattice. If site $1$ is empty, then a
particle is injected into this site with probability $\alpha$. If a particle is
present at site $L$ then it exits the system with probability
 $\beta$. By varying
$\alpha$ and $\beta$, the following phase diagram is obtained in steady state.
For $\alpha < \beta $ and $\alpha < \frac{1}{2}$, the system is in a
 low-density (LD) phase
 with a bulk density $\alpha$ and current $J = \alpha (1-\alpha)$. For 
$ \alpha > \beta$ and $\beta < \frac{1}{2} $  the system is in a high density
(HD) phase when the bulk properties are controlled by the exit conditions. In
this phase one has bulk density $1-\beta$ and current $J = \beta
(1-\beta)$. For $\alpha = \beta < \frac{1}{2}$, the system is in a coexistence
phase when part of the system has a bulk density $\alpha$ and
the remaining part has bulk density $(1-\beta)$. These two
parts are connected by a domain wall or shock. The $\alpha = \beta$ line is
known as the coexistence line which terminates at $(\alpha,\beta) =
(\frac{1}{2},\frac{1}{2}) $. 
For $\alpha , \beta > \frac{1}{2}$, the system is in a maximal
current (MC) phase with bulk density $\frac{1}{2}$, and current $J
=\frac{1}{4}$. This phase diagram was derived exactly in \cite{domany}. 
\begin{figure}
\includegraphics[scale=0.5]{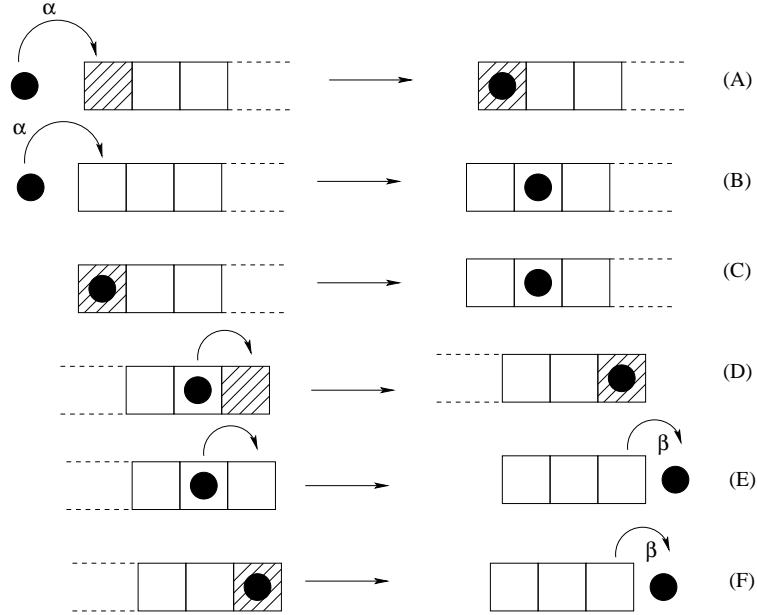}
\caption{Schematic representation of the moves close to the boundaries. The
sites are shown as boxes and particles as black solid circles. A
closed gate is shown as a shaded box at the boundary. 
 (A): Left gate is empty and in the closed state, particle injected with 
probability
$\alpha$ remains trapped at site $1$. (B): Left gate is empty and in the open 
state, particle injected with  probability
 $\alpha$ immediately detaches from site $1$ and hops to
site $2$. (C): Left gate is closed and occupied, switches to an open state
with probability $p_a$ and the particle is  immediately untrapped and hops to
site $2$. (D): Right gate is closed and empty, particle hopping from site
$L-1$ to $L$, gets trapped at site $L$. (E): Right gate is open and empty, particle
hops from site $L-1$ to $L$ and after hopping
immediately dissociates from site $L$ and exits the
system with probability $\beta$. (F): Right gate is closed and occupied, 
switches to an open state with probability $p_b$ when the trapped particle
 immediately dissociates and exits the system with probability $\beta$.}
\label{fig:model}
\end{figure}

In this paper, we consider a variant of the above system, where two stochastic
gates are present at the two boundary sites $1$ and $L$. The gating dynamics
is modeled so as to mimic ion-channel gating---these gates can
 stochastically switch
between high-affinity (closed) and low-affinity (open) states. The
configuration of the system is specified by specifying the occupancy of
each site and the state of the gates at the boundaries. If
the gate at site $1$ is in the closed or high-affinity state, then the
injected particle strongly binds to the site and therefore
remains trapped in that site. If the gate is in the open or
low-affinity state, then the particle immediately dissociates from site
$1$ and hops to the next site if it is empty. Similarly, if the gate at site
$L$ is closed, then a particle arriving at that site gets trapped. But if the
gate is open then the particle exits the system with probability $\beta$. 
Because of the hard core
constraint, no particle can enter the gate if it is already occupied. These
moves are shown schematically in Fig. \ref{fig:model}.
The switching rate of the gates are assumed to follow local detailed balance.
If $p_a$ is the probability that the left gate is open, then the ratio of the
forward and reverse transition rates between the open and closed states is
$(1-p_a)/p_a$. For the right gate the probability to be in open state is denoted 
by $p_b$ and the switching rates are chosen following similar local detailed 
balance. We are interested in how the $\alpha-\beta$ phase diagram is modified
for different choices of $p_a$ and $p_b$.

One important point to note here is that even in the limit $p_a=p_b=1$, {\sl
i.e.} when the gates are always in the open state, our model does not map onto
regular open-chain TASEP. Since the gating dynamics in our system is motivated
by the models for ion-channel gating \cite{dongen,andre11}, an open 
 state or low-affinity state of the gate implies
that any particle arriving at the gate immediately detaches and goes to 
the next site, as shown in Fig. \ref{fig:model}B \cite{dongen,andre11}. This
effect is not present for a regular TASEP where the waiting time at the
boundary sites is decided according to the rules of usual stochastic update.    
However, we have verified that all our qualitative conclusions remain valid
when we consider a variant of our model which maps onto regular TASEP in
$p_a=p_b=1$ limit. In this modified version, a particle gets trapped in a
closed gate as before, and
 it is allowed to leave while the gate is open, but it does
not immediately dissociate from an open gate. The waiting time of a particle
at the open 
gate is same as that at any other site. We have performed Monte Carlo
simulations and analytical calculations within mean-field theory in this model
and find similar results. See appendix \ref{sec:passive} for
more details.


\section{Monte Carlo simulations}
\label{sec:sim}

We perform Monte Carlo simulations of the above model. 
Starting from a random initial configuration, we evolve the system in time
following random sequential updates where we choose a site at random and update
it following the steps outlined in appendix \ref{sec:algo}. Each Monte Carlo
step consists of $L$ such update trials.  Since we are mainly interested in the
phase diagram, we use a large system-size $L=10^4$ in our
simulations so as to avoid errors due to finite size effects which
may cause significant difference close to the phase boundaries. The current 
and density profiles which characterize the phase of the system are measured 
after evolving for a long enough time to ensure that the system reaches 
steady state.    
\begin{figure}
\includegraphics[scale=0.6]{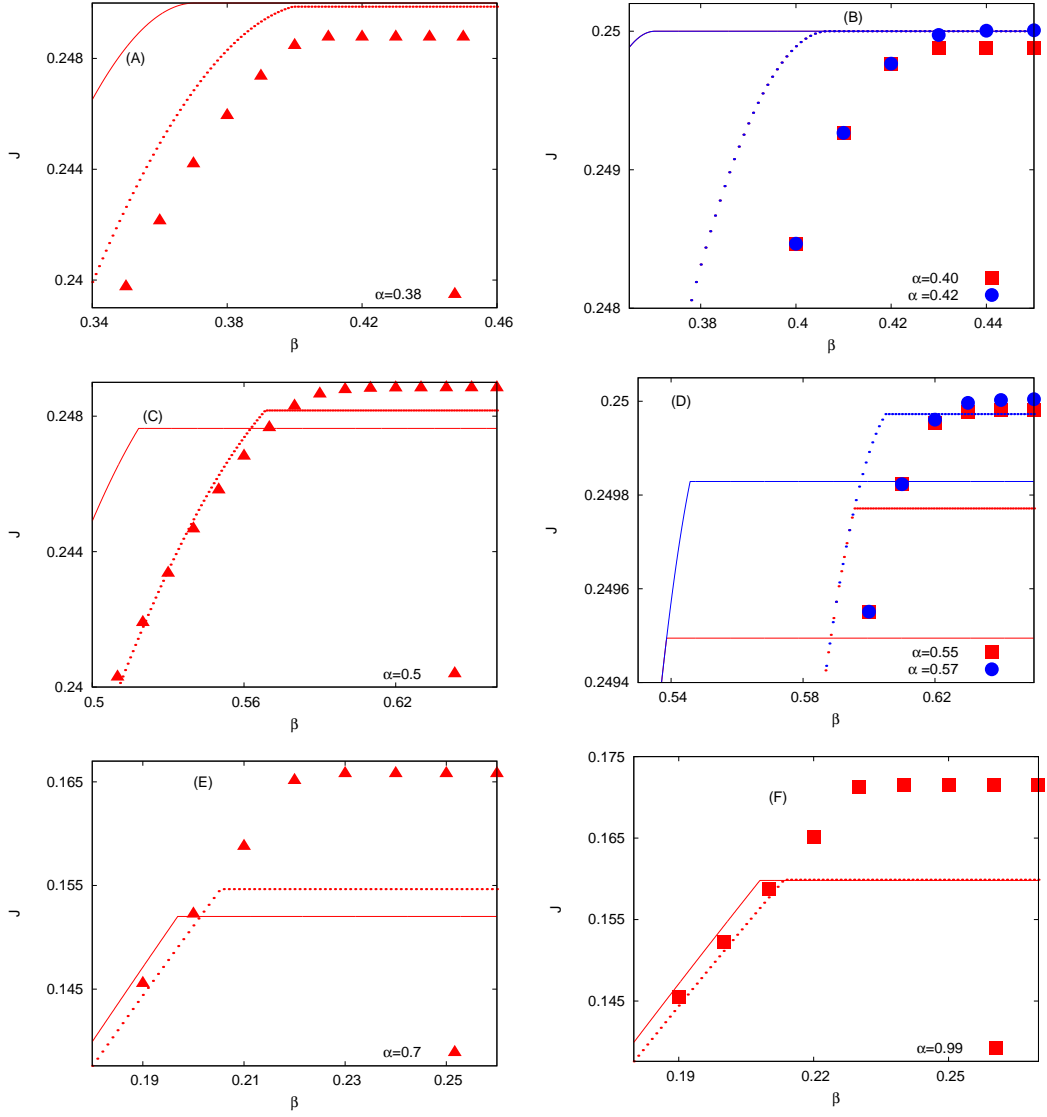}
\caption{Steady state current $J$ vs exit probability
 $\beta$ for fixed values of $\alpha$ and $p_a, p_b$. Discrete points show
simulation results, solid (dotted) continuous lines show analytical predictions from
mean-field (refined mean-field) theory.   
Top panel: $p_a=0.9$, $p_b=0.9$. (A): HD-LD transition at $\alpha =0.38$,
$\beta_c = 0.41 $ (simulation) and $0.40$ (refined mean-field). Mean-field
predicts HD-MC transition at $\beta_c = 0.37$. 
(B): simulation shows that for $\alpha = 0.4$ ($0.42$), HD-LD (HD-MC)
 transition at $\beta_c = 0.43 $ ($0.44$). 
 For both these $\alpha$ values, mean-field (refined
mean-field) theory predicts HD-MC transition at $\beta_c = 0.37$ ($0.41$). 
 Middle panel: $p_a=0.6$,  $p_b=0.6$. (C):
HD-LD transition for $\alpha = 0.5$ with $\beta_c = 0.59 $ (simulation),
$0.52 $ (mean-field) and $0.57 $ (refined mean-field). 
(D): HD-LD transition at  $\alpha = 0.55$ and $\beta_c = 0.63$
(simulation), $0.54$ (mean-field), $0.59$ (refined mean-field). For
 $\alpha = 0.57$ simulations show an HD-MC transition at $\beta_c = 0.63 $.
But mean-field and refined mean-field predict HD-LD transition at
$\beta_c=0.55$ and $0.6$, respectively. 
Bottom panel: $p_a=0.2$,  $p_b=0.8$. (E): HD-LD transition at $\alpha
= 0.7$, $\beta_c = 0.23 $ (simulation), $0.197$ (mean-field), $0.205 $
(refined mean-field). (F): HD-LD transition at
 $\alpha = 0.99$, $\beta_c = 0.24 $ (simulation), $0.208$ (mean-field) and
$0.213$ (refined mean-field). 
}
\label{fig:jb}
\end{figure}

To map out the phase diagram, we study the variation of the
 steady state current $J$ as a function of $\beta$ for fixed $\alpha$. 
The current is obtained by measuring the average flux across a bond per unit
time, averaged over a time-period of $4 \times 10^8$ Monte Carlo steps. The
measured current has an error-bar  $\simeq 5 \times 10^{-6}$ caused by
statistical fluctuations. 
 We find that for small $\beta$ when the system is in HD phase,
$J$ increases with $\beta$ and then saturates beyond a particular
$\beta_c$. If the saturation value of $J$ depends on $\alpha$, then $\beta_c$
marks the transition from HD to LD phase. If the saturation current takes an
$\alpha$-independent value $\frac{1}{4}$ then $\beta_c$ marks the transition
between HD and MC phase. Using this method, we calculate the phase-diagram for
various $p_a$ and $p_b$, from the plot of $J$ as a function of $\beta$ for
 fixed $\alpha$. In Fig. \ref{fig:jb} we show the data for $J$ vs $\beta$ for 
few representative values of $\alpha$ and $p_a$, $p_b$.   

\begin{figure}
\includegraphics[scale=0.9,angle=0]{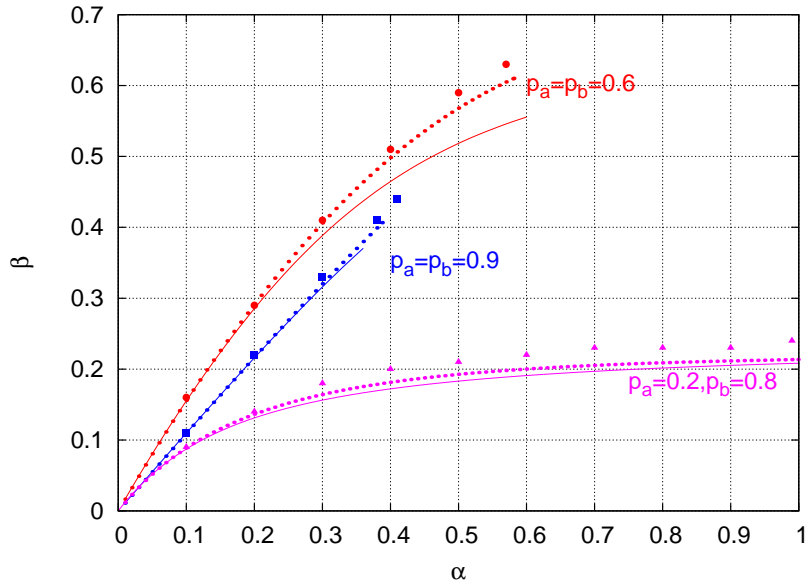}
\caption{Coexistence line between the HD and LD phase shown in $\alpha-\beta$
 plane for different values of $p_a$ and $p_b$. Continuous lines and dotted
lines show analytical predictions from simple mean-field theory and refined
mean-field theory, respectively. The simulation results are shown by discrete
points. The refined mean-field approach significantly improves the
quantitative agreement with simulation.}
\label{fig:coex}
\end{figure}

From the $J$ vs $\beta$ data we obtain $\alpha$ vs $\beta_c$ coexistence line 
for each given set of $p_a$ and $p_b$. In Fig. \ref{fig:coex} we plot these
coexistence lines. The critical values of $\alpha$ and $\beta$ that describe
this phase boundary have an error-bar of $0.01$ which comes from the
resolution of our measurement. For $p_a=p_b=0.9$ and $p_a=p_b=0.6$ the
coexistence line terminates at a particular point and the MC phase begins. 
But for $p_a =0.2$
and $p_b =0.8$, there is no MC phase, and for any $\alpha$ in the entire range
$0 \le \alpha \le 1$, one can find a $\beta_c$ that marks the HD-LD phase 
transition. Hence
for this case, the coexistence line extends all the way upto $\alpha =1$. 
In the following section, we derive these results analytically using a
 mean-field theoretic approach.

\section{Mean-Field theory}
\label{sec:mft}

Let $n_i$ be the occupancy of the $i$-th site in a
particular configuration. Within simple mean-field theory we assume all 
correlations
of the type $\langle n_i n_j \rangle$ are factorized and can be written as $\rho_i
\rho_j $, where $\rho_i = \langle n_i \rangle $ is the average density at the
$i$-th site. The time-evolution equations for these densities are
\begin{eqnarray}
\frac{\partial \rho_1}{\partial t} &=& \alpha (1-p_a)(1-\rho_1) +\alpha p_a
(1-\rho_1) \rho_2 - p_a \rho_1(1-\rho_2)
\label{eq:rho1}  \\
\frac{\partial \rho_2}{\partial t} &=& p_a \rho_1 (1-\rho_2) + \alpha p_a
(1-\rho_1) (1-\rho_2) - \rho_2 (1-\rho_3) \\
\frac{\partial \rho_i}{\partial t} &=& \rho_{i-1} (1-\rho_i) - \rho_i
(1-\rho_{i+1}) \;\;\;\; 3 \leq i \leq L-1 \\
\frac{\partial \rho_{L}}{\partial t} &=& \rho_{L-1}(1-\rho_L) (1-\beta p_b)
-p_b \beta \rho_L 
\label{eq:rhol}
\end{eqnarray}
Here, the first two terms on the right-hand-side of Eq. \ref{eq:rho1} represent
the possible mechanisms that lead to the increase of the density at the first 
site---the first term corresponds to the case when the left gate is closed and
 an injected particle gets trapped in site $1$ (as shown in Fig. \ref{fig:model}A),
 the second term stands for the
case when the left gate is open but the injected particle cannot immediately
detach from site $1$ and hop to site
$2$ because site $2$ is already occupied. The last term in that equation
describes the situation shown in Fig. \ref{fig:model}C, when the left gate
switches from closed to open state and the trapped particle is released and
goes to site $2$. In this way, all the terms in the above set of equations can
be explained. Although we have a total of four control parameters $\alpha,
\beta , p_a $ and $p_b$, from the time-evolution equations it follows that
$\beta$ and $p_b$ always appear in a product. So we treat $\beta p_b$ as a
single independent parameter.

Note that, as mentioned earlier, even when $p_a = p_b=1$, these equations are not identical to those
in regular open-chain TASEP. Consider the example 
 shown in Fig. \ref{fig:model}B, the site $2$
can directly receive a particle from the reservoir when the left gate is open.
Thus in the time-evolution equation for $\rho_2$ there is a term that couples
to $\alpha$. No such term is present for regular open-chain TASEP. Similarly,
at the right boundary, the time-evolution equation for $\rho_L$ has in general
a different form even when $p_b=1$. But here one can retrieve the
limit for regular TASEP by redefining the exit probability as $\beta \rightarrow \beta p_b /(1-\beta p_b)$.

In the steady state, all time derivatives on the left-hand-side of the
 equations (1)-(4) become zero and a constant current $J$ flows through 
every bond in the lattice. At the left end of the system, the injection
current, {\sl i.e.} the rate of influx from the reservoir to site $1$ is  
$J = \alpha (1-\rho_1)$. 
Depending on the state of the gate at site $1$ and on the occupancy of site
$2$, a fraction of the injected particles stays at site $1$ and the remaining
fraction immediately dissociates from site $1$ and hops to site $2$. Thus, the 
current flowing through the bond
between site $1$ and $2$ is given by $J = \alpha 
p_a (1-\rho_1) (1-\rho_2) + p_a (1-\rho_2)$. 
 We can now construct the following set of steady state equations, by
recognizing that the current $J$ has the same value through all the bonds of 
the system in the steady state.
\begin{eqnarray}
\rho_1 & = & 1-\frac{J}{\alpha} \\
\rho_2 &= & 1- \frac{J}{p_a (1+J - \frac{J}{\alpha})} \label{eq:rho2} \\
\rho_i & = & 1- \frac{J}{\rho_{i-1}}  \;\;\;\;\; 3 \leq i \leq L
\label{eq:rec} \\
\rho_L  & = & \frac{J}{\beta p_b} - J .
\label{eq:rhob}
\end{eqnarray}
Note that for $ 3 \leq i \leq L$, the local density $\rho_i$ satisfies the 
recursive relation in Eq. \ref{eq:rec}. 
Let the two fixed points of the recursive map be given 
by $\rho_\pm$, defined as $\rho_\pm = (1 \pm \sqrt{1-4J})/2$. It is easy to
see that $\rho_+$ is the stable fixed point and $\rho_-$ is the unstable fixed
point of the map. Also, $\rho_+ + \rho_-=1$ and $\rho_+ \rho_- = J$. Below we
identify different phases and the phase-boundaries following the same steps as 
in \cite{domany}.

\subsection{LD Phase} In this phase, the bulk density is infinitesimally
close to the unstable fixed point $\rho_-$ and deviates from this value close
to the right boundary, while satisfying $\rho_L < \rho_+$. Note that due to
presence of the stochastic gate at site $1$, the starting point of the 
recursive map is $\rho_2$, instead of $\rho_1$, unlike ordinary TASEP.  
 Putting $\rho_2 =\rho_- $ yields  
\begin{equation}
\rho_- = 1- \frac{J}{p_a(\rho_1 + J)} = 1-\frac{J}{p_a (J+1-\frac{J}{\alpha})} 
\end{equation} 
Using the relation between $\rho_+$ and $\rho_-$, one gets  
a quadratic equation for $\rho_-$:
\begin{equation}
\rho_- ^2 (1-\frac{1}{\alpha}) + \rho_- (\frac{1}{p_a}
+\frac{1}{\alpha}-1)-1=0
\end{equation} 
This has the solution
\begin{equation}
\rho_- = \frac{1}{2} + \frac{\alpha}{2 p_a (1-\alpha)} \mp \frac{\alpha}{2
(1-\alpha)} \left [ (3-\frac{1}{\alpha})^2 + (\frac{1}{p_a} -1)^2
+\frac{2}{\alpha p_a} -5 \right ]^{1/2} 
\label{eq:ld}
\end{equation}
Now, by definition, $\rho_- < 1/2$, and since the second term in the above
equation is always positive for $0 < \alpha <1 $, one must take the 
negative root and in addition one must satisfy
\begin{eqnarray}
\frac{1}{p_a} < \left [ (3-\frac{1}{\alpha})^2 + (\frac{1}{p_a} -1)^2
+\frac{2}{\alpha p_a} -5 \right ]^{1/2} 
\end{eqnarray}
which after simplification gives the condition  
\begin{equation}
(\alpha -1) (5 \alpha -\frac{2 \alpha}{p_a} -1) >0.
\label{eq:ldpa}
\end{equation}
Since $0 < \alpha <1$, above inequality is satisfied if  
\begin{eqnarray}
\alpha < \frac{1}{5-\frac{2}{p_a}} =\alpha _m, \;\;\;\; 5-\frac{2}{p_a} >0
\label{eq:am}
\end{eqnarray}
For  $(5-2/p_a)<0$, {\sl i.e.} for $p_a <2/5$, the inequality Eq.~(\ref{eq:ldpa}) 
 holds for all
 $\alpha$. As shown later, this implies that for $p_a <2/5$
 there is no MC phase in the system.

\subsection{HD phase}

In this phase, the density at the bulk and the right end 
is infinitesimally close to the stable fixed point $\rho_+$ and close to the
left boundary there is deviation from this value, with $\rho_2 > \rho_-$.
Using $\rho_L = \rho_+$ gives 
\begin{equation}
\rho_+ =1-\rho_- = \frac{1- 2 \beta p_b}{1-\beta p_b}
\label{eq:hd}
\end{equation} 
The condition $\rho_+ > 1/2$ implies
\begin{equation}
\beta  < \frac{1}{3p_b} = \beta_m
\label{eq:bm}
\end{equation}

\subsection{MC phase}
\label{sec:mc}
In this phase, $\rho_2 \ge 1/2$ and $\rho_L \le 1/2$. The current $J$ takes 
the maximal
 value $J=1/4$ and the bulk density is $1/2$. From conditions specified by Eqs.~(\ref{eq:rho2}) and (\ref{eq:rhob}) this phase can be observed if
\begin{equation} 
\alpha > \alpha_m =\frac{1}{5-\frac{2}{p_a}}; \;\; \; \; \; \beta >\beta_m 
=\frac{1}{3 p_b} 
\end{equation}
Since both $\alpha$ and $\beta$ must be less than unity, MC phase can be
observed if and only if
 $p_a >1/2$ and $p_b >1/3$. For $p_a$ or $p_b$ lower than these
limiting values MC phase does not exist. Note that this takes care of
 the condition $p_a > 2/5$ derived in Eq. \ref{eq:am}.

\subsection{HD-LD Coexistence line}
On the critical line separating HD
and LD phase, we match the expression for $\rho_- = 1-\rho_+$ obtained from Eqs.
(\ref{eq:ld}) and (\ref{eq:hd}) for the two phases:
\begin{equation}
  \frac{1}{2} + \frac{\alpha}{2 p_a (1-\alpha)} - \frac{\alpha}{2
(1-\alpha)} \left [ (3-\frac{1}{\alpha})^2 + (\frac{1}{p_a} -1)^2
+\frac{2}{\alpha p_a} -5 \right ]^{1/2} = \frac{\beta p_b}{1-\beta p_b}
\end{equation} 
This is the equation for the coexistence line which terminates at $\alpha =
\alpha_m$ and $\beta = \beta_m$ and MC phase starts from this point. 
It follows from this equation that when the gates are always open, {\sl i.e.}
$p_a = p_b =1$, then the coexistence line is given by $\alpha = \beta$, as in
regular TASEP, but it terminates at $\alpha_m = \beta_m= 1/3$, instead of 
$\alpha_m = \beta_m= 1/2$.
 However, for general values of $p_a$ and $p_b$, the coexistence
line has a curvature.

We compare the coexistence line obtained from
mean-field theory (continuous lines) with that from simulation (discrete closed
symbols) in Fig. \ref{fig:coex}. We find that there is good qualitative
agreement between the two but there is a systematic quantitative deviation. For
example, at $p_a=0.2$, $p_b =0.8$, our simulation shows there is no MC phase
and this is consistent with the criterion derived in section \ref{sec:mc}.
However, mean-field systematically under-estimates $\beta_c$ values for all
$\alpha$ and the mismatch becomes more prominent for large $\alpha$ values.
This also affects the MC phase boundaries. For $p_a=p_b=0.9$, mean-field
predicts that the coexistence line should terminate at $\alpha_m =
\frac{9}{25} =0.36$ and
$\beta_m = \frac{10}{27} \simeq 0.37$. But our simulation gives $\alpha_m = 0.41$ and $\beta_m
= 0.44$. Similarly, for $p_a=p_b=0.6$, mean-field prediction is $\alpha_m =
\frac{3}{5} = 0.6$ and $\beta_m = \frac{5}{9} \simeq 0.56$, whereas our
simulation yields $\alpha_m = 0.57$ and $\beta_m = 0.63$. Thus there is
significant mismatch between the mean-field and simulation results.
  To improve the quantitative agreement with
simulations, we develop a refined mean-field theory in the next section.

\section{Refined Mean Field Theory}
\label{sec:rmft}

Unlike the simple mean field theory described above,
 where all correlations between the sites are
ignored, in this refined mean-field approach, we take into account certain
correlations in the system, which are expected to be more important than
others, in a self-consistent way, as was first demonstrated in a model for
a dynamically extending exclusion process \cite{evans3}. 
In the bulk of the system one
still expects the correlations to be weak. But close to the boundary this
assumption breaks down and strong correlations may be present between the sites.
 For example, in the limit of small $p_a$ when the
residence time of the left gate in the closed state is large, an injected
particle remains trapped at site $1$ for long enough such that the
particle at site $2$ (if any) has time to drift away towards the bulk of the
system before the gate opens. When the gate opens finally, the trapped
particle hops to site $2$. Therefore, for small values of $p_a$, a strong 
anti-correlation develops between sites $1$ and $2$. For general values of
$p_a, \alpha$ and $\beta$, depending on the interplay of the time-scales
associated with boundary injection, exit and switching process of the left
gate, the correlation between the sites $1$ and $2$ can become positive or
negative and vary in strength. Similarly, at the right boundary a correlation
develops between sites $L-1$ and $L$.

Let $C_a = \langle n_1 n_2 \rangle$ and $C_b = \langle n_{L-1} n_L \rangle $
be the correlation between the sites near the left and right boundary,
respectively. Unlike
simple mean-field theory, we do not replace these correlations by simple
product, rather we treat $C_a$ and $C_b$ as additional variables. All other
correlations in the system are assumed to be factorized. Therefore, the
equations of motion are: 
 
\begin{eqnarray}
\frac{\partial \rho_1}{\partial t} &=& \alpha (1-p_a)(1-\rho_1) + \alpha p_a
(\rho_2 -C_{a}) - p_a (\rho_1 - C_{a})
\label{eq:r1} \\
\frac{\partial \rho_2}{\partial t} &=& p_a (\rho_1-C_{a}) + \alpha p_a
(1-\rho_1 -\rho_2 + C_{a}) - \rho_2 (1-\rho_3) \\
\frac{\partial C_{a}}{\partial t} &=& \alpha (\rho_2-C_{a}) - C_{a}
(1-\rho_3)
\label{eq:ca} \\ 
\frac{\partial \rho_i}{\partial t} &=& \rho_{i-1} (1-\rho_i) - \rho_i
(1-\rho_{i+1}) \;\;\;\;\; 3 \leq i \leq L-2 \\
\frac{\partial \rho_{L-1}}{\partial t} &=& \rho_{L-2}(1-\rho_{L-1}) -
\rho_{L-1} +C_b 
\label{eq:rl-1} \\
\frac{\partial \rho_{L}}{\partial t} &=& (\rho_{L-1}-C_b) (1-\beta p_b)
-p_b \beta \rho_L
\label{eq:rl} \\
\frac{\partial C_{b}}{\partial t} &=& \rho_{L-2}(\rho_L -C_b) - \beta p_b C_b
\label{eq:cb}
\end{eqnarray}
In steady state, when all time-derivatives vanish, we express $C_a$, $\rho_1$,
$\rho_3$ in Eqs \ref{eq:r1} and \ref{eq:ca} in terms of $\rho_2$ and $J$ to
obtain the following equation
\begin{equation}
\alpha p_a \rho_2 ^2 + \alpha \rho_2 \left ( J - p_a + \frac{p_a J}{\alpha}
\right ) + J \left [ J (1-p_a) - p_a \left (1-\frac{J}{\alpha} \right ) \right
] =0
\label{eq:r2j}
\end{equation}
For a given $J$ the above quadratic equation can be solved for $\rho_2$:
\begin{equation}
\rho_2 = \frac{1}{2} -\frac{J}{2} \left ( \frac{1}{p_a} + \frac{1}{\alpha}
\right ) \pm
\frac{1}{2 \alpha p_a} \sqrt{\alpha ^2 \left ( J - p_a + \frac{p_a J}{\alpha}
\right )^2 - 4 \alpha p_a J \left [ J(1-p_a) - p_a \left ( 1-\frac{J}{\alpha}
\right ) \right ]}
\label{eq:r2q}
\end{equation}

From here, one can work out the conditions for the MC phase, where $J = 1/4$
and $\rho_2 \ge 1/2$. The last criterion implies that in Eq. \ref{eq:r2q}
only positive root should be taken and one must have
\begin{equation}
\alpha ^2 \left ( J - p_a + \frac{p_a J}{\alpha} \right )^2 - 4
\alpha p_a J \left [ J(1-p_a) - p_a \left ( 1-\frac{J}{\alpha}\right ) \right ] \ge (\alpha J + p_a J)^2
\end{equation}   
For $J=1/4$ in MC phase this condition gives 
\begin{equation}
\alpha ^2 (p_a ^2 - \frac{p_a}{2}) + \frac{\alpha}{4} (3 p_a ^2 - p_a)
-\frac{p_a^2}{4} \ge 0
\label{eq:mca}
\end{equation}
This is the condition that must be satisfied at the left boundary 
of the chain to
observe MC phase. To work out the condition for the right boundary, we consider
Eqs. \ref{eq:rl-1}, \ref{eq:rl} and \ref{eq:cb} from where we replace
$\rho_L$, $\rho_{L-2}$ and $C_b$ in terms of $\rho_{L-1}$ and $J$ to obtain
the following equation.
\begin{equation}
\beta^2 p_b^2 \rho_{L-1}^2 - \beta p_b (J + \beta p_b + \beta p_b J
)\rho_{L-1} + J (J+\beta p_b) =0 
\label{eq:rlj}
\end{equation} 
For a given $J$ this equation can be solved for $\rho_{L-1}$ 
\begin{equation}
\rho_{L-1} = \frac{1}{2} + \frac{J}{2} \left ( 1+ \frac{1}{\beta p_b} \right ) 
\pm \frac{1}{2 \beta p_b} \sqrt{ (J + \beta p_b + \beta p_b J )^2 - 4 J (J +
\beta^2 p_b^2) }
\end{equation}
from which negative root must be taken since in MC phase $\rho_{L-1} \le 1/2$.
In addition, one must satisfy
\begin{equation}
\frac{1}{2 \beta p_b} \sqrt{ (J + \beta p_b + \beta p_b J )^2 - 4 J (J +
\beta^2 p_b^2) } \ge  \frac{J}{2} \left ( 1+ \frac{1}{\beta p_b} \right )
\end{equation}
and for $J=1/4$ this gives
\begin{equation}
\beta \ge \frac{ \sqrt{3} -1}{2 p_b}
\label{eq:mcb}
\end{equation}
Eqs. \ref{eq:mca} and \ref{eq:mcb} define the conditions that must be
satisfied in order for the system to support an MC phase.

The HD-LD coexistence line can be obtained from matching the solutions of Eqs.
\ref{eq:r2j} and \ref{eq:rlj}. For example, in Eq. \ref{eq:rlj} we use the
conditions for HD phase and substitute
$\rho_{L-1} = \rho_+$ and $J = \rho_+ (1-\rho_+)$ to obtain
\begin{equation}
\rho_+ ^2 + \rho_+ (\beta ^2 p_b ^2 + \beta p_b -2) +1-\beta p_b -\beta ^2 p_b ^2 =0
\end{equation}
which has a solution
\begin{equation}
\rho_+ = 1-\beta p_b -\beta ^2 p_b ^2 
\label{eq:rp}
\end{equation}
and another trivial solution $\rho_+ =1$ which we neglect. Since $\rho_+ >
1/2$, one must have $\beta p_b (1+\beta p_b) < 1/2$ for HD phase, which is
consistent with the criterion obtained in Eq. \ref{eq:mcb}. 

For LD phase we use Eq. \ref{eq:r2j} and substitute $\rho_2 = \rho_-$ and $J =
\rho_- (1-\rho_-)$  to obtain the following cubic equation
\begin{equation}
\rho_-^3 \left ( 1-p_a + \frac{p_a}{\alpha} \right ) + \rho_-^2 \left (p_a
-\frac{2p_a}{\alpha} -\alpha -2 \right ) + \rho_- \left ( \alpha p_a + \alpha
+ p_a + \frac{p_a}{\alpha} +1 \right ) - (\alpha p_a + p_a) =0
\label{eq:cube}
\end{equation}
The solution for $\rho_-$ obtained from this equation must match that obtained
in Eq. \ref{eq:rp} on the HD-LD coexistence line. Therefore, the equation for
coexistence line is obtained by substituting $\rho_-=\beta p_b (1+\beta p_b)$
in  Eq. \ref{eq:cube}. 
\begin{figure}
\begin{center}
\includegraphics[scale=0.35]{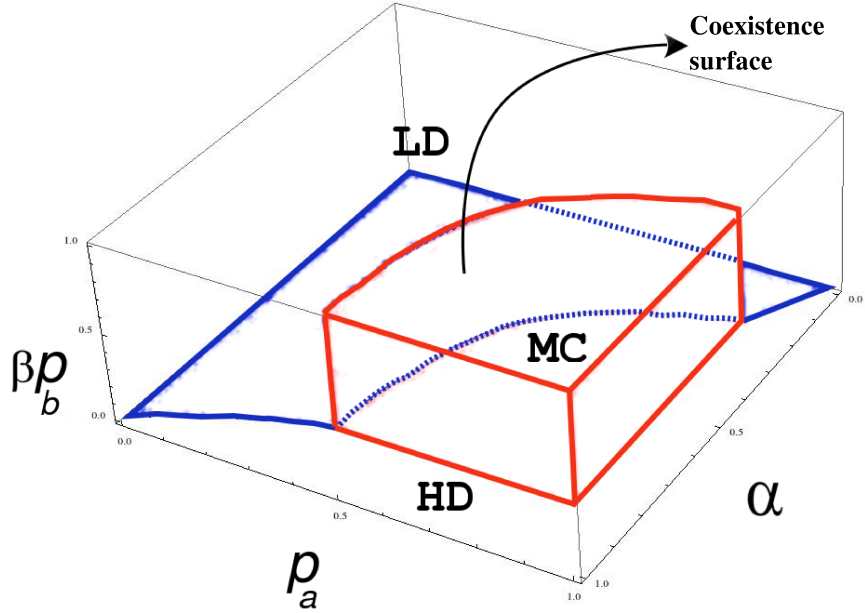}
\end{center}
\caption{Complete phase diagram in three dimensions obtained from refined 
mean-field theory with control parameters $\alpha$, $p_a$ and $\beta p_b$.
The coexistence surface is the one demarcated by the thick blue lines. The
region enclosed by the thick red lines indicates the maximal current phase.
The region above the coexistence surface is the low-density phase, while 
the region below the coexistence surface is the high-density phase.}
\label{fig:3d}
\end{figure}

In Fig. \ref{fig:jb} we show variation of $J$ with
$\beta$ (dotted lines) for different $p_a, p_b$ values. The critical point 
$\beta_c$ is the value of $\beta$ for which the current $J$ reaches a 
saturation. As can be seen from the figure, refined mean field theory 
consistently predicts a more accurate value for $\beta_c$ as compared to mean 
field theory. The actual value of the current is also predicted better by 
the refined mean field theory. The remaining discrepancy between the theory
and simulation results may be attributed to the presence of non-zero 
correlations extending into the bulk, beyond the first and last site 
correlations taken into account in our refined mean-field approach. 
In Fig. \ref{fig:coex} we plot the coexistence line (dotted line) from 
refined mean-field theory  and compare with the simulation results.
Even in this case we find that compared
to simple mean-field, the refined mean-field prediction shows a better
quantitative agreement with simulation results. In Fig. \ref{fig:3d} 
we present the complete 
phase diagram in three dimension with the three independent parameters, 
$\alpha$, $p_a$ and $\beta p_b$ as axes. The coexistence region 
(Eq.~\ref{eq:cube}) is a surface in this three dimensional space, as shown in 
Fig. ~\ref{fig:3d}.
  
\section{Conclusion}
\label{sec:discussion}

In this paper, we have studied the phase diagram of an open-chain TASEP whose
entrance and exit at the boundaries are controlled by two gates, whose
dynamics are chosen so as to mimic ion-channel gates in cell membrane.
In our model, these gates
can switch between a high-affinity state, when it strongly binds to the
particles thereby trapping them, and a low-affinity state, when particle
immediately dissociates from the gates and move to the next site. This
immediate dissociation implies even in the limit when the gates are always open,
our model does not map onto a regular open-chain TASEP. However, all our
results and conclusions remain valid in a slightly
different version of the model, where there is no immediate dissociation in
the open state (see appendix \ref{sec:passive}) and usual TASEP limit is
recovered, when the gates are always open.

The effect of stochastic gates on the phase diagram was studied earlier where
a two-way coupling between the gating dynamics and particle flow was
considered \cite{wood}. While entrance of exit of particles can take place
only through an open gate, immediately after passage the gate shuts close. On
the other hand, in our model, no such two-way coupling is assumed and the
gates can switch between two states, completely independent of the particle
flow. Even this uncoupled, independent gating dynamics yields important
modification in the phase diagram which cannot be reproduced by mere rescaling
of entrance and exit rates.

Stochastic gating, as investigated in our model, allows systems to actively 
control the phase diagram, and hence the current flow through the lattice, 
independent of the entry and exit rates. The entry and exit rates often depend
on the concentration of particles near the boundary sites, 
and gating then allows
a mechanism for systems to prevent flow even in the presence of high
 concentrations 
(or equivalently, high entry rate) and vice versa. This opens
new possibilities for mechanisms for the regulation of particle current, 
especially in biological systems. It might be of interest to include more
 biologically relevant features in our model and check how generic our present 
results are.

\section*{Acknowledgments}
 MKM would like to acknowledge support from the Ramanujan Fellowship
(RJN-09/2012), Department of Science and Technology, India and the IRCC Seed
Grant from IIT-Bombay.


\appendix
\renewcommand{\theequation}{A-\arabic{equation}}
\setcounter{equation}{0}
\renewcommand{\thefigure}{A-\arabic{figure}}
\setcounter{figure}{0}

\section{Simulation Algorithm}
\label{sec:algo}

We evolve the system following random sequential update where a randomly
chosen site is updated according to the steps outlined below. 

(a) Choose an integer $k$ within $[0,L]$.

(b) If $1<k<L-1$, then a particle hops from site $k$ to $k+1$,
 if the $k$-th site is occupied and $(k+1)$-th site is empty.

(c) If $k=L-1$, and if $(L-1)$-th site is occupied and $L$-th site is empty, a
particle hops from $L-1$ to $L$. After the hopping takes place, the particle
 attempts to exit the system  with probability $\beta$, if 
the gate at site $L$ is open. If the hopping attempt is unsuccessful, the
 particle stays at site $L$.

(d) If $k=0$, then if site $1$ is empty, a particle is injected into site $1$
with probability $\alpha$. If the gate at site $1$ is open, the injected 
particle immediately hops to site $2$, if site $2$ is empty.  

(e) If k=1, and if the gate at site $1$ is open, close it with probability
 $1-p_a$. If the gate is closed, open it with probability
 $p_a$  and if there is a particle, it hops to site $2$, if site $2$ is empty.

(f) If $k=L$ and if the gate at site $L$ is open, close it with probability
 $1-p_b$. If the gate is closed, open it with probability $p_b$ and
 after opening the gate, the particle at site $L$, if any, attempts to 
exit the system with probability $\beta$.


\renewcommand{\theequation}{B-\arabic{equation}}
\setcounter{equation}{0}
\renewcommand{\thefigure}{B-\arabic{figure}}
\setcounter{figure}{0}

\section{Open state without immediate dissociation}
\label{sec:passive}

In this section, we discuss a variant of our model that maps onto the regular
open-chain TASEP in the limit when the gates are always open. The only
difference between this model and the one discussed in the main part of the
paper is that in the open state of the gate there is no immediate
dissociation. In other words, the hopping rate of a particle from 
an open gate is same as the hopping rate from any bulk site (as opposed to an
infinite hopping rate that stands for immediate dissociation). The mean-field
equations for this system read
\begin{eqnarray}
\frac{\partial \rho_1}{\partial t} &=& \alpha (1-\rho_1) - p_a \rho_1(1-\rho_2) 
 \\
\frac{\partial \rho_2}{\partial t} &=& p_a \rho_1 (1-\rho_2) - \rho_2 (1-\rho_3)
 \\
\frac{\partial \rho_i}{\partial t} &=& \rho_{i-1} (1-\rho_i) - \rho_i
(1-\rho_{i+1}) \;\;\;\; 3 \leq i \leq L-1 \\
\frac{\partial \rho_{L}}{\partial t} &=& \rho_{L-1}(1-\rho_L) 
-\beta p_b  \rho_L 
\end{eqnarray} 
It is easy to see that in the limit $p_a=p_b=1$ this model maps onto regular
open-chain TASEP. 

Following the method illustrated in section \ref{sec:mft} these equations can
be solved to derive the following equation for the coexistence line
\begin{equation}
1+\frac{\alpha}{p_a} - \left [\left ( 1+\frac{\alpha}{p_a} \right )^2 - 4
\alpha \right ]^{1/2} = 2 \beta p_b 
\end{equation}  
which terminates at $\alpha_m = \dfrac{1}{2(2-1/p_a)}$ and $\beta_m =
\dfrac{1}{2 p_b}$ and MC phase starts from here. The coexistence line obtained
from simulation (data not shown) matches qualitatively with this calculation
but there is a quantitative mismatch. This is taken care of when refined
mean-field theory is used. Following the steps outlined in section
\ref{sec:rmft}, it is easy to show that within refined mean-field
theory the equation for the coexistence line becomes
\begin{equation}
\alpha \beta^2 p_b^2 -\beta p_b (1+\alpha -\beta p_b) \left [1-\beta p_b (1-\beta p_b)\left (
\frac{1}{\alpha} + \frac{1}{p_a} \right ) \right ] =0
\end{equation} 
This line shows much better quantitative agreement with the simulation (data
not presented here).


%

\begin{thebibliography}{99}

\bibitem{gunter} G.M. Sch{\"u}tz and K.J. Wiese, {\it Phase transitions and
critical phenomena} {\bf 19} 1 (2001).

\bibitem{menon} Y. Aghababaie, G.I. Menon and M. Plischke, {\it Phys. Rev. E}
{\bf 59} 2578 (1999)

\bibitem{lipowsky} R. Lipowsky, S. Klumpp and T.M. Nieuwenhuizen, {\it Phys.
Rev. Lett.} {\bf 87} 108101 (2001).

\bibitem{klumpp} S. Klumpp and R. Lipowsky, {\it J. Stat. Phys.} {\bf 113} 233
(2003).

\bibitem{evans1} K.E.P. Sugden, M.R. Evans, W.C.K. Poon and N.D. Read, {\it
Phys. Rev. E} {\bf 75} 1909 (2007).

\bibitem{evans2} M.R. Evans and K.E.P. Sugden {\it Physica A} {\bf 384} 53 (2007)

\bibitem{evans3} K.E.P. Sugden and M.R. Evans, {\it J. Stat. Mech.}
P11013 (2007).

\bibitem{schreck} D. Chowdhury, D.E. Woif and M. Schreckenberg, {\it Physica A}
{\bf 235} 417 (1997).

\bibitem{schads} D. Chowdhury, L. Santen and A. Schadschneider {\it Phys.
Rep.} {\bf 329} 199 (2000)

\bibitem{helbing} D. Helbing, {\it Reviews of Modern Physics} {\bf 73} 1067
(2001).

\bibitem{hilhorst} H.J. Hilhorst and C. Appert-Rolland, {\it J. Stat. Mech.}
P06009 (2012).

\bibitem{krug} J. Krug, {\it Phys. Rev. Lett.} {\bf 67} 1882 (1991).

\bibitem{henkel} M. Henkel and G. Sch{\"u}tz, {\it Physica A} 
{\bf 206} 187 (1994).

\bibitem{domany} B. Derrida, E. Domany and D. Mukamel, {\it J. Stat. Phys.}
{\bf 69} 667 (1992).

\bibitem{derrida} B. Derrida, M.R. Evans, V. Hakim and V. Pasquier, {\it J.
Phys. A: Math. Gen.} {\bf 26} 1493 (1993).

\bibitem{gunter93} G. Sch{\"u}tz and E. Domany, {\it J. Stat. Phys.}
 {\bf 72} 277 (1993).

\bibitem{zia} T. Chou, K. Mallick and R.K.P. Zia, {\it Rep. Prog. Phys.} {bf
74} 116601 (2011).

\bibitem{chowdhury} D. Chowdhury, A. Schadschneider and K. Nishinari, {\it
Phys. Life Rev.} {\bf 2} 318 (2005).

\bibitem{wood} A.J. Wood, {\it J. Phys. A: Math. Theor.} {\bf 42}, 
445002 (2009).


\bibitem{hille} Hille, {\it B. Ionic Channels of Excitable Membranes}, Sinauer, Sunderland, MA (1992).

\bibitem{nehar} E. Neher and B. Sakmann, Nature 260, 799–802 (1976).

\bibitem{hodgkin} A. L. Hodgkin and A. F. Huxley, {\it J. Physiol.} {\bf 117}, 500 (1952).

\bibitem{dongen} A.M.J. VanDongen, {\it Proc. Nat. Acad. Sc.} {\bf 101}, 3248
(2004).

\bibitem{andre11} D. Andreucci, D. Bellaveglia, E.N.M. Cirillo, and S.
Marconi,  {\it Phys. Rev. E} {\bf 84}, 021920 (2011).

\end{thebibliography}
\end{document}